\documentclass[a4paper]{article}
\usepackage{graphicx}

\def\d {\mbox{d}}
\begin{document}
\title{Relativistic analysis of an earth-satellite time transfer }

\author{O. Minazzoli \& B. Chauvineau \footnote{Observatoire de la C\^ote d'Azur, ARTEMIS, CNRS UMR 6162, 06130 Grasse, France}} 

\maketitle

\begin{abstract}
Analytical treatment of time transfer problem for Earth-Satellite system is presented. The development was made in a complete relativistic framework. In accordance with modern clock precision and for low altitude orbits, we neglect the other bodies and consider only the $\frac{1}{c^2}$ Earth potential developed up to the $J_2$ term in spherical harmonics.\end{abstract}
%
\section{Introduction}

During the last few years, new technological developments increase significantly the time transfer accuracy using laser links. Further technological developments will go beyond the precision of the currently used relativistic theoretical description generally used in data reduction programs. At present time, only the first order in the post newtonian metric is considered and the Earth potential is reduced to the monopolar term, which corresponds to a spherical Earth (Blanchet \& al. 2001). 

In foreseeable future laser links between ground based stations and low orbit satellites (T2L2, ACES) would reach precisions of order $10^{-14}$ or even $10^{-15}$ s. In this context, we show modelizing relativistic effects considering spherical Earth only is not sufficient. Hence, we consider the other effects to select the ones which have to be taken into account at this level of accuracy.

In sec. 2, we derive the orders of magnitude of the different relativistic effects, related to both Earth's potential and external bodies. We show, only the Earth's $J_2$ potential term has to be included. In sec. 3, we give the analytical expressions of the trajectory and the corresponding propagation time.

\section{Selecting the relevant terms for low orbit satellites}

In general relativity, photons follow null geodesics of the space-time
metric $g_{\mu \nu }$, which encodes the gravitational interaction.
Following the IAU2000 convention (Soffel \& al. 2003), we take the following form for the metric
(one uses units such that $G=c=1$) 

\begin{equation}
\label{eq:metric}
ds^{2}=g_{\mu \nu }dx^{\mu }dx^{\nu }\equiv -(1-2U)dt^{2}+(1+2U)\delta
_{ij}dx^{i}dx^{j}.
\end{equation}
This metric is developed up to first order in the potential\ $U$, which
represents the sum of the newtonian form of the potentials associated to
each gravitational source.\ This form of the metric will turn out to be
sufficient for the proposed application. The sources of gravitation are the
Earth and external bodies. As usual, greek indices ($\alpha,\beta,$...) run from 0 to 3 (space-time variables), while latin indices ($i,j,k,$...) run from 1 to 3 (spatial variables).

The Earth potential can be developed in spherical harmonics.\ The monopolar
term (spherical Earth) writes 
\[
U_{E,m}=%
\frac{M_{E}}{r}
\]
where $M_{E}$ is Earth's mass.\ Beyond the monopolar term, the dominant term
is the so called $J_{2}$-term, which, in brief, corresponds to the Earth's
oblateness contribution to the gravitational field. This term is of the
order of 
\[
U_{E,J_{2}}\sim J_{2}\frac{M_{E}R_{E}^{2}}{r^{3}}
\]
where $R_{E}$\ is Earth's radius. The other terms in the multipolar
development are at least two orders of magnitude weaker. The (tidal)
contribution of each external body (essentially the Moon and the Sun) to the
potential is of the order of 
\[
U_{ext}\sim x^{i}x^{j}\partial _{i}\partial _{j}U\sim \left( \frac{r}{L}%
\right) ^{2}\frac{M}{L}
\]
where $M$ and $L$\ are the mass and the distance of the involved body.

$U_{E,m}$ has relative effects of the order of $10^{-9}$ on the motion of a photon.\
Hence, for a $\sim 1000\;km$ altitude satellite, the effect on the flying
time of a photon linking a ground based station and this satellite can reach 
$\sim 10^{-11}$ or some $10^{-12}\;s$. Since $J_{2}$ is of order $10^{-3}$,
the contribution of this term is of order $10^{-14}$ or some $10^{-15}\;s$.
The contribution of the other multipolar terms is then at best of order $%
10^{-16}\;s$. It turns out that the contributions of external bodies are of
the order of the $J_{2}$ contribution for altitudes corresponding to
geostationary satellites. Besides, for $r<17000\;km$, it would be meaningless taking external
perturbations into account if harmonic terms beyond $J_{2}$\ are not
included in the Earth's potential model. Second order terms
in the metric (neglected in the metric presented) are of order $10^{-18}$,
hence induce time delay corrections of order $10^{-20}\;s$.

The precision which could be reached by time transfer experiments in a foreseeable future is of
the order of $10^{-15}\;s$.\ Hence, considering satellites at altitudes $\sim
1000\;km$, it is necessary to include, beyond the Earth's monopolar term,
the Earth's $J_{2}$ contribution to the potential, but it is legitimate to
neglect all the other contributions.

\section{Obtaining the time transfer}

As usual in relativistic solar system experiments, we take the following form of the metric (constants G and c have been explicitly included):

\begin{eqnarray}
ds^2=g_{\mu \nu} dx^{\mu} dx^{\nu} \equiv -\left(1-\frac{2U}{c^2}\right) c^2 dt^2 + \left(1+\gamma \frac{2U}{c^2}\right) \mid d\vec{r} \mid^2\\ \nonumber
U= \frac{G M_E}{r}+ J_2 \frac{G M_E}{2r} \left( \frac{R_E}{r} \right)^2 \left( 1- 3 \left(\frac{z^2}{r^2} \right) \right)
\end{eqnarray}
which generalizes the (\ref{eq:metric}) metric, in order to include viable alternative geometric gravity theories, like scalar-tensor theories (Will 1993). The deviation of those theories to general relativity is encoded in the $\gamma$ term which is unity in general relativity. Current experimental tests and astronomical observations show $\mid \gamma-1 \mid$ is less than some $10^{-5}$ (Will 2006). Hence, a possible deviation from general relativity should not have measurable effects in our problem. Since making computations in this more general context does not lead to dramatic complications, we present results in the general case, but it is always possible to set $\gamma=1$ if the general relativity case is considered.
The light geodesic equation writes :

\begin{equation}
\frac{\d k_{\alpha}}{\d \lambda}= \frac{1}{2} k^\mu k^\nu \partial_\alpha g_{\mu \nu}
\end{equation}
with $k_{\alpha}=g_{\alpha \beta} k^\beta$, $k^\alpha=\frac{\d x^\alpha}{\d \lambda}$, $\lambda$ being an affine parameter along the geodesic. The wave vector $k^\alpha$ is isotropic which means $k^\alpha k_\alpha = 0$. At first order in the potential, the geodesic equation leads to:

\begin{eqnarray}
\frac{\stackrel{(\rightarrow 1)}{\d k_0}}{\d \lambda}=0\\
\frac{\stackrel{(\rightarrow 1)}{\d k_i}}{\d \lambda}=(1+\gamma) \frac{G M_E}{c^2} \left[\partial_i \frac{1}{r} + \frac{J_2}{2} {R_E}^2 \partial_i \frac{1}{r^3} - \frac{3}{2} J_2 {R_E}^2 \partial_i \frac{z^2}{r^5} \right]
\end{eqnarray}
where $\stackrel{(\rightarrow 1)}{Q^\alpha} \equiv~ \stackrel{(0)}{Q^\alpha}+\stackrel{(1/2)}{Q^\alpha}+\stackrel{(1)}{Q^\alpha}$ is the development of $Q^\alpha$ up to the first order in $U$ of any quantity $Q^\alpha$. Up to first order in $U$ the solution reads :
\begin{equation}
\stackrel{(\rightarrow 1)}{x^0}=x^0_0 + f(\vec{x},\vec{n}) \lambda + c^{-1} g(\vec{x},\vec{n})
\end{equation}
\begin{center}
$f(\vec{x},\vec{n}) = \left( 1+ 2 \frac{G M_E}{c^2} J_2 R_E^2 \left\{ A \frac{1}{r} + B \frac{1}{r^3} \right\} \right)
$\\
$
g(\vec{x},\vec{n}) = 2 \frac{G M_E}{c^2} \left[ ln\left( \frac{r+\vec{n}\cdot \vec{x}}{r_0+\vec{n}\cdot \vec{x_0}} \right) + J_2 R_E^2 \left\{ A ~~\vec{n} \cdot \vec{x_0} \left(\frac{1}{r}-\frac{1}{r_0}\right) + B ~~\vec{n} \cdot \vec{x_0}  \left(\frac{1}{r^3}-\frac{1}{{r_0}^3} \right) + C~~\left(\frac{1}{r^3}-\frac{1}{{r_0}^3} \right) \right\} \right]
$
\end{center}
$A,B,C$ are constant along the geodesic, $x^\alpha_0$ correspond to initial conditions and $\vec{n}$ is a normalized tri-vector ($\vec{n} \cdot \vec{n}=1$) related to the components of the wave vector $k^\alpha$. We get from this formulae the function $\stackrel{(\rightarrow 1)}{\lambda} (x^0,\vec{n})$, from which we obtain $\stackrel{(\rightarrow 1)}{x^i}\left(\stackrel{(\rightarrow 1)}{\lambda}(x^0,\vec{n})\right)$ and find (writing, as usual, $x^0=t$):
\begin{eqnarray}
\label{eq:xi}
\stackrel{(\rightarrow 1)}{x^i}(t,\vec{n})=x^i_0 +n^i c (t-t_0) +\frac{G M_E}{c^2} f^i(t-t_0,\vec{n})\\ \nonumber
f^i(t-t_0,\vec{n})=(1+\gamma) \left[ -n^i ln \left(\frac{r+\vec{x} \cdot \vec{n}}{r_0+\vec{x_0} \cdot \vec{n}} \right) - B^i (r-r_0) + J_2 R_E^2~~ h^i(t-t_0, \vec{n}) \right] \\ \nonumber
h^i(t-t_0, \vec{n})=\Omega^i \left( \frac{t-t_0}{r^3} - \vec{n} \cdot \vec{x_0} \left(\frac{1}{r^3} - \frac{1}{r_0^3} \right) \right)
+\Delta^i \left( \frac{t-t_0}{r} - \vec{n} \cdot \vec{x_0} \left(\frac{1}{r} - \frac{1}{r_0} \right) \right)
\\ \nonumber+\Xi^i \left(\frac{1}{r^3} - \frac{1}{r_0^3} \right)
+\Theta^i \left(\frac{1}{r} - \frac{1}{r_0} \right) 
+\Gamma^i (r-r_0 )
\end{eqnarray}
with $B^i,\Omega^i,\Delta^i,\Xi^i,\Theta^i,\Gamma^i$ constant along the geodesic. Obtaining the time transfer is then solving the equation $\stackrel{(\rightarrow 1)}{x^i}(t,\vec{n})=\stackrel{(\rightarrow 1)}{x_{sat}^i}(t)$, where $\stackrel{(\rightarrow 1)}{x_{sat}^i}(t) \equiv x^i_{s 0}+v^i_s(t_0) (t-t_0)+ \frac{a^i_s}{2}(t_0) (t-t_0)^2$. We get $\stackrel{(\rightarrow 1)}{T} \equiv \stackrel{(0)}{T}+\stackrel{(1/2)}{T}+\stackrel{(1)}{T}$ with $T=t-t_0$ :

\begin{eqnarray}
\stackrel{(0)}{T}=c^{-1} \sqrt{\left( \vec{x_{s 0}} - \vec{x_0} \right)^2}\\
\stackrel{(1/2)}{T}= \left(\stackrel{(0)}{\vec{n}} \cdot \frac{\vec{v_s}}{c}\right) \stackrel{(0)}{T}\\
\stackrel{(1)}{T}= \frac{1}{2} \left[ \left( \frac{\vec{v_s}}{c} \right)^2 + \left(\stackrel{(0)}{\vec{n}} \cdot \frac{\vec{v_s}}{c}\right)^2 + \left(\stackrel{(0)}{\vec{n}} \cdot \frac{\vec{a_s}}{c} \right) \stackrel{(0)}{T} \right] \stackrel{(0)}{T} - \frac{G M_E}{c^3} \left( \stackrel{(0)}{\vec{n}} \cdot \vec{f}\left(\stackrel{(0)}{T},\stackrel{(0)}{\vec{n}}  \right) \right)
\end{eqnarray}
with $\stackrel{(0)}{n^i}=\frac{x^i_{s 0}-x^i_0}{\sqrt{\left( \vec{x_{s 0}} - \vec{x_0} \right)^2}}$.

\section{Conclusion}
As a consequence of the loss of spherical symmetry, a lot of new terms
appears from the $J_2$ potential term ( see \ref{eq:xi}). Possible next steps in order to upgrade our analysis could involve the next order in the metric development ($g_{0 i} = O(3/2)$) and/or the next terms in the spherical harmonic development ($J_3,C_{22},...)$. Thus, to go further, the next development could be based on a 3/2 order metric with the earth potential developed up to the $J_2$ term or more if needed. External potentials are not relevant for low altitude orbits and should be neglected in future studies.

\end{document}